# Transverse optical binding for a dual dipolar dielectric nanoparticle dimer

Xiao-Yong Duan,[1,2*] Graham D. Bruce,[2] Kishan Dholakia,[2,3] Zhi-Guo Wang,[4] Feng Li,[1] and Ya-Ping Yang[5]

[1] *School of Mathematics, Physics and Information Engineering, Jiaxing University, Jiaxing 314001, China*
[2] *SUPA School of Physics and Astronomy, University of St Andrews, St Andrews, KY16 9SS, UK*
[3] *Department of Physics, College of Science, Yonsei University, Seoul 03722, South Korea*
[4] *School of Physics Science and Engineering, Tongji University, Shanghai 200092, China*
[5] *School of Design, Jiaxing University, Jiaxing 314001, China*

*Corresponding author: xyduan@zjxu.edu.cn

**Abstract:** The physical origins of the transverse optical binding force and torque beyond the Rayleigh approximation have not been clearly expressed to date. Here, we present analytical expressions of the force and torque for a dual dipolar dielectric dimer illuminated by a plane wave propagating perpendicularly to the dimer axis. Using this analytical model, we explore the roles of the hybridized electric dipolar, magnetic dipolar, and electric-magnetic dipolar coupling interactions in the total force and torque on the particles. We find significant departures from the predictions of the Rayleigh approximation, particularly for high-refractive-index particles, where the force is governed by the magnetic interaction. This results in an enhancement of the dimer stability by one to four orders of magnitude compared to the predictions of the Rayleigh approximation. For the case of torque, this is dominated by the coupling interaction and increases by an order of magnitude. Our results will help to guide future experimental work in optical binding of high-refractive-index dielectric particles.

## I. INTRODUCTION

Optical binding originating from the mutual scattering of electromagnetic (EM) waves among objects enables micro- and nanoparticles to form stable spatial arrangements. Therefore, it has attracted great attention in light-controlled self-assembly of particles [1, 2]. For a pair of particles (or dimer), when the incident light wave vector is *perpendicular* to the dimer axis, the resulting force between particles is named the transverse optical binding force (TOBF) [3]. For the case of the light wave propagating *along* the dimer axis, the force is termed the longitudinal optical binding force [4].

TOBF offers a novel way to assemble dielectric and metallic particles into complex steady structures, such as one-dimensional chains [3] and waveguides [5], two-dimensional optical matter [6, 7], clusters [8], arrays [9-11], and mirrors [12], and three-dimensional clusters [13]. Furthermore, it can dynamically manipulate particles, e.g. inducing oscillations in particle chains [14], rotation and spinning of particles [15, 16], and forming a bound state of two rotating micro-gyroscopes [17]. Currently, an emerging research field is the dependence of the TOBF on EM hybridization coming from the interaction between electric and/or magnetic multipoles in adjacent particles. In detail, the sign of the TOBF in a silver disk-ring nanostructure reverses multiply due to the hybridized interference between the electric dipole of disk and the electric high-order modes of ring [18]. Analogous phenomena were found in Au nanorod heterodimers [19]. Further numerical results showed that the TOBF for a silicon dimer changes from an attractive force to a repulsive one with variation of wavelength [20]. The repulsive force originates from the hybridization between the broad electric dipole and narrow magnetic dipole in the particles. Recent theory indicated that TOBF induced by the interference of surface plasmon polarizations (SPPs) of two metallic particles over a metallic substrate is an order of magnitude larger than the TOBF without SPPs [21].

While the TOBF is well-studied for low-refractive-index dielectric (index<2.5, e.g. polystyrene and glass) and metallic particles in the Rayleigh approximation [15, 22], optical binding for high-refractive-index dielectric particles with $n$>3.5, such as silicon and germanium, has received little attention because analytical solutions of the TOBF beyond the Rayleigh approximation have remained absent [1, 23, 24]. Such particles are of interest in the area of dielectric metamaterials, where they can be harnessed for optical devices without the high heating rates associated with plasmonic particles [25], and in optical trapping because their strong oscillatory dynamics can be utilized as a thermal engine [26]. Further, their morphology can be manipulated to customize their form birefringence to provide optical microfluidic actuators [27], and they can be used to create nano-heterostructure semiconductor devices [28]. In this paper, we analytically study the TOBF and torque on two dual dipolar dielectric particles orthogonally-illuminated by an arbitrarily polarized plane EM wave, as shown in Fig. 1. We show that the contributions to the force and torque due to electric dipolar, to magnetic dipolar, and to electric-magnetic dipolar coupling interactions are dominant in different regimes of refractive index, and therefore are essential for the full description of the TOBF, going beyond the Rayleigh approximation.

The paper is organized as follows. In Sec. II, we present the results of our derivation of analytical expressions for the TOBF and optical torque. Meanwhile, the physical meanings of the expressions are discussed. In Sec. III, the contributions of the electric dipolar, magnetic dipolar, and electric-magnetic dipolar coupling interactions to TOBF, torque, and stability of the dimer are numerically investigated in detail. Finally, conclusions are drawn in Sec. IV.

## II. THEORETICAL MODEL

Based on the optical force on a single dual dipolar particle [29] and EM mutual scattering between two particles [30] (see details in Appendixes A-F), we derive the TOBF ($F$) (exerted along the *y*- axis in Fig. 1) on particle B, which is equal but oppositely directed to the force on particle A. $F$ consists of three parts:

$$F_e = \frac{\varepsilon_0 \varepsilon_s}{2} \left\{ \begin{array}{l} \left( \Re\left[\frac{\partial \kappa}{\partial R}\right] |\tilde{\alpha}_{ey}|^2 - \Re\left[\frac{\partial \mu}{\partial R}\right] |\tilde{\alpha}_{em-p,1}|^2 + \Re\left[\frac{\partial \eta}{\partial R} \tilde{\alpha}_{em-p,2} \tilde{\alpha}^*_{em-p,1}\right] \right) |E_{0,y}|^2 \\ + \left( \Re\left[\frac{\partial \eta}{\partial R} \tilde{\alpha}_{em-s,1} \tilde{\alpha}^*_{em-s,2}\right] + \Re\left[\frac{\partial \mu}{\partial R}\right] |\tilde{\alpha}_{em-s,2}|^2 \right) |E_{0,x}|^2 \end{array} \right\},$$
(1)

$$F_m = \frac{\varepsilon_0 \varepsilon_s}{2} \left\{ \begin{array}{l} \left( \Re\left[\frac{\partial \mu}{\partial R}\right] |\tilde{\alpha}_{em-p,2}|^2 + \Re\left[\frac{\partial \eta}{\partial R} \tilde{\alpha}_{em-p,1} \tilde{\alpha}^*_{em-p,2}\right] \right) |E_{0,y}|^2 \\ + \left( \Re\left[\frac{\partial \kappa}{\partial R}\right] |\tilde{\alpha}_{my}|^2 - \Re\left[\frac{\partial \mu}{\partial R}\right] |\tilde{\alpha}_{em-s,1}|^2 + \Re\left[\frac{\partial \eta}{\partial R} \tilde{\alpha}_{em-s,2} \tilde{\alpha}^*_{em-s,1}\right] \right) |E_{0,x}|^2 \end{array} \right\},$$
(2)

$$F_{em} = \frac{\varepsilon_0 \varepsilon_s}{12\pi} k^4 \Im\left[ \tilde{\alpha}_{em-p,1} \tilde{\alpha}^*_{em-p,2} |E_{0,y}|^2 - \tilde{\alpha}^*_{em-s,1} \tilde{\alpha}_{em-s,2} |E_{0,x}|^2 \right],$$
(3)

where $\Re$ and $\Im$ represent the real and imaginary parts of a complex number, * denotes complex conjugation, $\varepsilon_0$ and $\varepsilon_s$ are respectively the permittivity of vacuum and relative permittivity of the medium, $E_{0,x}$ and $E_{0,y}$ are $x$ and $y$ components of the incident electric field, $R$ is the distance between the centers of the two particles, $\partial/\partial R$ denotes the partial derivative with respect to the distance, $k=2\pi/\lambda_s$ is the wavenumber in the medium, and the dressed polarizibilities $\tilde{\alpha}_{ey}$, $\tilde{\alpha}_{my}$, $\tilde{\alpha}_{em-p,1}$, $\tilde{\alpha}_{em-p,2}$, $\tilde{\alpha}_{em-s,1}$ and $\tilde{\alpha}_{em-s,2}$ represent the polarization of the particle induced by the incident light and hybridization while $\mu$, $\kappa$, and $\eta$ are eigenvalues of electric and magnetic dyadic Green's functions of a point dipole (see details in Appendixes A and B).

$F_e$ is the electric component of $F$. The first term denotes the interaction between the $y$-components of the electric dipole moments ($p_y$) of the two particles, while the second term is the interaction between the $z$-components of the electric dipole moments ($p_z$). The third term is the force on the $z$-component of the electric dipole moment in particle B ($p_z^B$) acted on by the $x$-component of the magnetic dipole moment in particle A ($m_x^A$). The combination of these first three terms (in the first parenthesis) represents $F_e$ in the case of illumination by $p$-polarization (i.e. along the dimer axis). The fourth term is the interaction between the two $p_x$ while the last one denotes the force exerted on $p_x^B$ by $m_z^A$. These last two terms (in the second parenthesis) represent $F_e$ for illumination by $s$-polarization. On the other hand, $F_m$ is the magnetic component of $F$. The first term denotes the interactions between two $m_x$ while the second one is the force exerted on $m_x^B$ by $p_z^A$. The first two terms (in the first parenthesis) represent $F_m$ in case of $p$-polarization. The third and fourth terms denotes respectively the interactions between two $m_y$ and between two $m_z$. The last term shows the force exerted on $m_z^B$ by $p_x^A$. The last three terms (in the second parenthesis) determine $F_m$ for $s$-polarization. Finally, $F_{em}$ is the electric-magnetic coupling component of $F$. The first term comes from the interference between $p_z^B$ and $m_x^B$ and represents $F_{em}$ for $p$-polarization. The last term originates from the interference between $p_x^B$ and $m_z^B$ and represents $F_{em}$ in the case of $s$-polarization.

For Rayleigh particles ($\alpha_m=0$) whose refractive index or radius is small enough to satisfy the relation $kr=2\pi n_p r/\lambda \ll 1$, $F$ is determined by only the first and fifth terms in Eq. (1) while $F_m = F_{em} = 0$. In this case, the force is the classical optical binding force between two electric dipolar particles, as reported in Ref. [15]. On the other hand for magnetic dipolar particles ($\alpha_e=0$) such as Au core-Si shell nanosphere [31], the only non-zero contributions to $F$ are the first and third terms in Eq. (2), which together constitute the magnetic binding force [32], while $F_e = F_{em} = 0$.

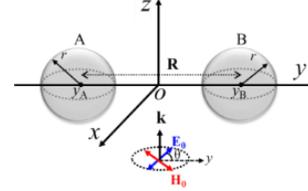

Fig. 1. A pair of dielectric spheres with radius $r$ located symmetrically at positions $y_A$ and $y_B$ along the $y$ axis of a Cartesian system ($O$-$xyz$). $R$ represents the distance between the two centers of particles. A plane EM wave with electric ($\mathbf{E}_0$) and magnetic ($\mathbf{H}_0$) field vectors is incident on the dimer along the $z$ axis, as shown by the wave vector $\mathbf{k}$. $\theta$ is the polarization angle of the wave, measured between the $\mathbf{E}_0$ and $y$ axis.

In addition to the binding force, the incident light also causes an extrinsic optical torque ($\Gamma$) on the dimer [15] which causes the dimer to rotate around the $z$ axis. The torque is also composed of three parts:

$$\Gamma_e = \frac{\varepsilon_0 \varepsilon_s}{2} \left\{ \begin{array}{l} 2\Re[\kappa - \mu] \Re\left[ \tilde{\alpha}_{em-s,2} \tilde{\alpha}^*_{ey} E_{0,x} E^*_{0,y} \right] \\ + \Re\left[ \eta \left( \tilde{\alpha}_{my} \tilde{\alpha}^*_{em-p,1} - \tilde{\alpha}_{em-s,1} \tilde{\alpha}^*_{ey} \right) E_{0,x} E^*_{0,y} \right] \end{array} \right\},$$
(4)

$$\Gamma_m = \frac{\varepsilon_0 \varepsilon_s}{2} \left\{ \begin{array}{l} -2\Re[\kappa - \mu] \Re\left[ \tilde{\alpha}_{em-p,2} \tilde{\alpha}^*_{my} E^*_{0,x} E_{0,y} \right] \\ + \Re\left[ \eta \left( \tilde{\alpha}_{em-p,1} \tilde{\alpha}^*_{my} - \tilde{\alpha}_{ey} \tilde{\alpha}^*_{em-s,1} \right) E^*_{0,x} E_{0,y} \right] \end{array} \right\},$$
(5)

$$\Gamma_{em} = \frac{\varepsilon_0 \varepsilon_s k^4 R}{12\pi} \Im\left[ \left( \tilde{\alpha}_{ey} \tilde{\alpha}^*_{em-s,1} + \tilde{\alpha}_{em-p,1} \tilde{\alpha}^*_{my} \right) E^*_{0,x} E_{0,y} \right].$$
(6)

$\Gamma_e$ is the electric component of $\Gamma$. The first term denotes the interaction of $p_x^B$ with $p_y^A$ and the interaction of $p_y^B$ with $p_x^A$. The last two terms (in the last square brackets) indicate respectively the interaction of $p_z^B$ with $m_y^A$ and interaction of $p_y^B$ with $m_z^A$. In addition, $\Gamma_m$ is the magnetic component of $\Gamma$. The first term represents the interaction of $m_x^B$ with $m_y^A$ and interaction of $m_y^B$ with $m_x^A$. The last two (in the last square brackets) represent respectively the interaction of $m_y^B$ with $p_z^A$ and interaction of $m_z^B$ with $p_y^A$. Moreover, $\Gamma_{em}$ is the electric-magnetic coupling component of $\Gamma$. The first term originates from the interference between $p_y^B$ and $m_z^B$ while the second one comes from the interference between $p_z^B$ and $m_y^B$.

For Rayleigh particles, $\Gamma_e$ is determined by only the first term in Eq. (4) while $\tilde{\alpha}_{em-s,2}$ reduces to $\tilde{\alpha}_{ex}$. This agrees with the classical torque on a Rayleigh-approximation dimer [15]. Meanwhile $\Gamma_m=\Gamma_{em}=0$. On the other hand, for purely magnetic dipolar particles,

$\Gamma_m$ is determined only by the first term in Eq. (5) and $\tilde{\alpha}_{em-p,2}$ is simplified to $\tilde{\alpha}_{mx}$, in agreement with magnetic optical torque between two magnetic dipolar particles in [32]. In this case, $\Gamma_e=\Gamma_{em}=0$. Note that for $p$- and $s$-polarizations the torque vanishes. Equations (1)-(6) are general for a plane wave in the near-field, intermediate-field, and far-field. They are the main results in this paper. Our numerical calculations are based on Eqs. (1)-(6).

Furthermore, the force and torque can be simplified in the far-field region ($kR\gg1$) where we retain only the highest order terms of $kR$ in $\mu$, $\kappa$ and $\eta$. For $p$-polarization, Equations (1)-(3) are respectively simplified as

$$F_e = \frac{n_s I_0}{c}\left\{\frac{k^2|\alpha_e|^2}{2\pi R^2}\cos(kR) + \frac{k^5|\alpha_m|^2}{(4\pi R)^2}\Im[\alpha_e]\right\}, \quad (7)$$

$$F_m = -\frac{n_s I_0}{c}\cdot\frac{k^3|\alpha_m|^2}{4\pi R}\sin(kR), \quad (8)$$

$$F_{em} = \frac{n_s I_0}{c}\cdot\frac{k^6|\alpha_m|^2}{24\pi^2 R}\{\cos(kR)\Re[\alpha_e] - \sin(kR)\Im[\alpha_e]\}, \quad (9)$$

where $I_0=\varepsilon_0 n_s c|E_0|^2/2$ is the intensity of the incident wave in the medium, $c$ is light speed in vacuum, $E_0$ is the intensity of incident electric field, $\alpha_e=i6\pi a_1/k^3$ and $\alpha_m=i6\pi b_1/k^3$ are the electric and magnetic polarizabilities with radiation reaction terms of the particles where $a_1$ and $b_1$ are respectively the electric and magnetic dipolar Mie scattering coefficients [33], and $i$ is unit imaginary number. Equations (7)-(9) demonstrate that $F_e$ is proportional to $R^{-2}$ as shown by the red curve in Fig. 2 (a) while $F_m$ and $F_{em}$ are proportional to $R^{-1}$ as shown by the blue and green curves. The results indicate that $F_e$ decays faster than $F_m$ and $F_{em}$ in the far-field. For Rayleigh particles, the binding force is completely determined by the first term in Eq. (7) which is in agreement with Ref. [34]. On the other hand for a purely magnetic dipolar dimer, the binding force is completely determined by Eq. (8). In the case of $s$-polarization, $F_e$ is expressed by Eq. (8) by replacing $\alpha_m$ with $\alpha_e$, which is the binding force in the Rayleigh approximation. $F_m$ and $F_{em}$ are respectively written as Eqs. (7) and (9) by exchanging $\alpha_m$ and $\alpha_e$. Therefore, $F_m \propto R^{-2}$ as shown by the blue curve in Fig. 2 (b) while $F_e$ and $F_{em} \propto R^{-1}$ as shown by the red and green curves. It means $F_m$ decays faster than $F_e$ and $F_{em}$. The first term in $F_m$ represents the binding force between two pure magnetic dipolar particles. Finally, for left-hand circular polarization with $\mathbf{E}_0 = E_0(1,i,0)\exp(ikz)/\sqrt{2}$ [35], $F_e$ and $F_m$ are expressed by Eq. (8) through employing respectively $\alpha_e$ and $\alpha_m$. In this case, $F_e=F_{Ray}$ and $F_m$ is the binding force between two magnetic dipolar particles. Moreover, the coupling component of the binding force is read as

$$F_{em} = \frac{n_s I_0}{c}\cdot\frac{k^6}{48\pi^2 R}\begin{Bmatrix}\cos(kR)(|\alpha_m|^2\Re[\alpha_e]+|\alpha_e|^2\Re[\alpha_m]) \\ -\sin(kR)(|\alpha_m|^2\Im[\alpha_e]+|\alpha_e|^2\Im[\alpha_m])\end{Bmatrix}. \quad (10)$$

The above results show that the binding force $F \propto R^{-1}$. Additionally, the electric component of torque is read as

$$\Gamma_e = \frac{n_s I_0}{2c}\cdot\frac{k^4}{(4\pi R)^2}\begin{Bmatrix}(|\alpha_e|^2+|\alpha_m|^2)\Im[\alpha_e] \\ -|\alpha_e|^2\begin{pmatrix}\cos(2kR)\Im[\alpha_m-\alpha_e] \\ +\sin(2kR)\Re[\alpha_m-\alpha_e]\end{pmatrix}\end{Bmatrix}. \quad (11)$$

The magnetic component of torque $\Gamma_m$ is also expressed by Eq. (11) through exchanging $\alpha_e$ and $\alpha_m$. Moreover, the coupling component of torque $\Gamma_{em}$ is written as

$$\Gamma_{em} = \frac{n_s I_0}{2c}\cdot\frac{k^6}{24\pi^2}\begin{Bmatrix}\cos(kR)(|\alpha_e|^2\Im[\alpha_m]+|\alpha_m|^2\Im[\alpha_e]) \\ +\sin(kR)(|\alpha_e|^2\Re[\alpha_m]+|\alpha_m|^2\Re[\alpha_e])\end{Bmatrix}. \quad (12)$$

Notice that $\Gamma_e$ and $\Gamma_m \propto R^{-2}$ while, interestingly, $\Gamma_{em}$ is independent of $R$. Therefore, the torque is dominated by the coupling interaction which is completely different to the binding force. Furthermore, the torque in Rayleigh approximation is simplified from Eq. (11) as

$$\Gamma_{Ray} = \frac{n_s I_0}{2c}\cdot\frac{k^4|\alpha_e|^2}{(4\pi R)^2}\{(\cos(2kR)+1)\Im[\alpha_e]+\sin(2kR)\Re[\alpha_e]\}. \quad (13)$$

On the other hand, for a magnetic dipolar dimer, the torque $\Gamma_{MD}$ is also expressed by Eq. (13) through replacing $\alpha_e$ by $\alpha_m$. Notice that we focus on the optical binding between dual dipolar particles without considering the electric quadrupole-assisted force. This is because the electric quadrupole is not dominant compared to the dipolar resonances within our parameters.

### III. RESULTS AND DISCUSSIONS
#### A. The TOBFs for $p$- and $s$-Polarized Waves

We consider two spheres with high refractive index $n_p=4$, i.e. Germanium [36], immersed in water ($n_s=1.33$) (A discussion of the forces and torques on particles with lower refractive index is given in Appendices G – L). The wavelengths of the incident wave in vacuum and water are $\lambda_0=532$nm and $\lambda_s=\lambda_0/n_s=400$nm, respectively. The power density of the wave in water is $I_0=10$mW/µm². To demonstrate the applicability of our model beyond the Rayleigh regime, we model spheres of radius $r=100$nm or 150nm. As expected, in the small $r$ limit, our model will degenerate to the Rayleigh approximation, as discussed in Eqs. (1)-(6). The force, torque, and distance ($R$) are respectively in units of pN, pN•µm, and $\lambda_s$.

Figure 2 shows the $F$ (black solid curve) as well as $F_e$ (red solid curve), $F_m$ (blue solid curve), and $F_{em}$ (green solid curve) between dielectric particles with $r=100$nm and $n_p=4$ as a function of the distance $R$ between the centers of the two particles for (a) $p$-polarized and (b) $s$-polarized waves. Additionally, $F_{Ray}$ (purple dashed curve) for a Rayleigh-approximation dimer ($\alpha_m=0$) is calculated by Eq. (3b) in Ref. [15] as a comparison. It can be seen from Fig. 2 (a) that in the case of $p$-polarization $F_m$ greatly exceeds $F_e$ and $F_{em}$ and dominates $F$. Physically, the effects of hybridization are particularly pronounced for high-refractive-index particles. For example, the re-radiated field by $m_x$ in one particle is strong enough to induce $p_z$ in the neighboring one. Thus $F_e$ does not only depend on the interaction between $P_y^A$ and $P_y^B$, which is the case for Rayleigh particles, but also by the interactions between $m_x^A$ and $p_z^B$ and between $p_z^A$ and $p_z^B$. Previous theoretical work demonstrated that the magnetic TOBF of an ideal magnetic dimer can be enhanced to the same magnitude as the TOBF for Rayleigh particles [32]. But what is surprising here is that the $F_m$ is an order of magnitude larger than not only $F_{Ray}$ but also $F_e$. The reason is that the $p_y$ induced in the two particles by the incident wave is nearly suppressed at wavelength $\lambda_0$=532nm while the two $m_x$ are excited even though they do not reach the maximum resonance (see Figs. 2 and 3 in Ref. [37]). As a result, $F$ is dominated by contributions from $F_m$ while the stable equilibrium positions of $F$ (black circles) are determined by those of $F_m$ (blue stars) instead of those of $F_e$ (red triangles) and $F_{em}$ (blue diamonds). The results demonstrate that $F_{Ray}$ dramatically underestimates $F$ for dielectric particles with high refractive index. Similar phenomena are observed in $s$-polarization in Fig. 2 (b). With increasing refractive index, the relative contribution of $F_e$ to the binding force is gradually overtaken by that due to $F_m$ (see details in Appendixes H and I). However, we emphasize that even for low refractive index particles such as polystyrene ($n_p$=1.59), non-negligible components of $F_m$ and $F_{em}$ already arise, and the magnitudes of these exceed the magnitude of $F_e$ for moderate refractive indices such as silicon ($n_p$=3.5) in $p$-polarization (see details in Appendix G and H). Additionally, $F$ in the case of $p$-polarization is larger than that for $s$-polarization. The reasons are that the interaction between two side-by-side parallel $m_x$ in $p$-polarization is larger than that between two head-to-tail collinear $m_y$ in $s$-polarization, while the electric dipolar interactions are suppressed.

Moreover, we calculated the first stable equilibrium positions of $F$, $F_e$, $F_m$, and $F_{em}$ (see details in Appendix I) as well as the stiffness at these positions for dielectric particles with $n_p$ from 1.4 to 4 in cases of $p$-and $s$-polarizations in Fig. 3. Within this range, many materials of interest in optical trapping are concentrated, e.g. polystyrene ($n_p$ =1.59) [38], diamond ($n_p$ =2.4) [39], silicon($n_p$ =3.5) [40], bismuth($n_p$ =3.89) [28], and germanium ($n_p$ =4) [26]. The particle's radius is set to 150nm in order to improve further its magnetic response because the magnetic dipole moment caused by the displacement current in the particle increases with the particle size. In detail, for $p$-polarization in Fig. 3 (a), the stiffness of $F_e$ (red curve with triangles) decreases sharply in the blue region, meaning that $F_e$ is barely sufficient for the formation of a stable bound state. The stable equilibrium vanishes completely at $n_p$=3.1 (vertical black dashed line) where $F_e$ has no stiffness and formation of a stable dimer is prohibited (see details in Appendix J). On the other hand, the stiffness of $F_m$ (blue curve with stars) is much larger than those of $F_{em}$ (green curve with diamonds) and $F_e$. Especially, for particles with $n_p$ from 2.7 to 3.6, the stiffness of $F_e$ decreases rapidly and even vanishes, because the suppressed electric dipolar responses cause the electric dipole-dipole interaction to provide barely-sufficient potential well depths to bind the two particles. However, the stiffness of $F_m$ is 1 to 4 orders of magnitude larger than that of $F_e$. Surprisingly, out of this region, the stiffness of $F_m$ is still one order of magnitude larger than that of $F_e$. This means that $F$ (black curve with circles) is mainly dominated by $F_m$, which shows the predominant ability of $F_m$ to bind the dielectric dimer for $p$-polarization. For $s$-polarization in Fig. 3 (b), the stiffness of $F_e$ and $F_{em}$ are of the same order of magnitude and much larger than that of $F_m$. Interestingly, in the region 2.9< $n_p$ <3.24 (faint blue area), the stability of the dimer is dominated by $F_m$; but in the neighboring region 3.24<$n_p$<3.5 (faint red area), the stability is dominated by $F_e$. Therefore, $F_e$ and $F_m$ supplement each other to enable stable formation of the dimer. It can be clearly seen from Fig. 3 that the magnetic dipolar interaction dominates the stability of the dimer in $p$-polarization while the electric dipolar interaction is dominant in $s$-polarization.

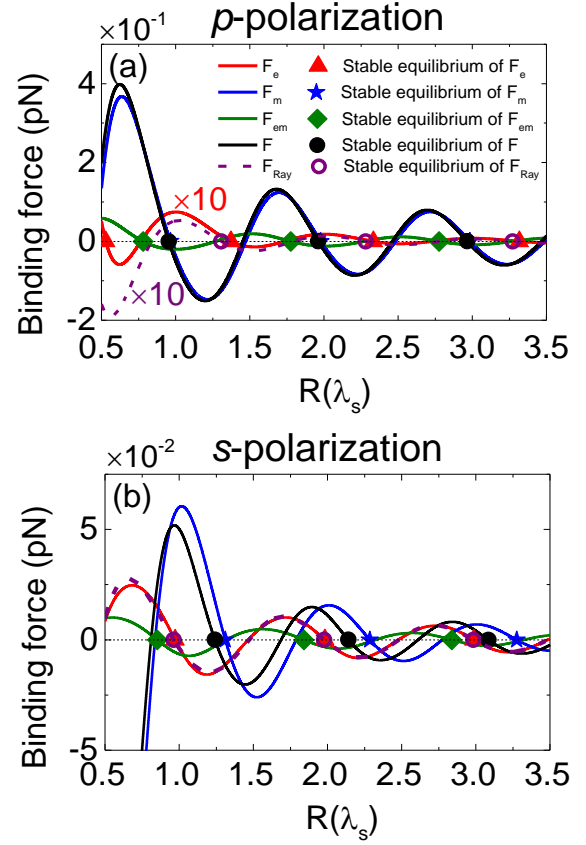

Fig. 2. Binding force versus distance ($R$) between the centers of two spheres ($r$=100nm and $n_p$=4) for $p$-polarization (a) and $s$-polarization (b). $F$ (black) is the binding force while $F_e$ (red), $F_m$ (blue), and $F_{em}$ (green) are respectively the electric, magnetic, and coupling components of $F$. $F_{Ray}$ (purple) is the binding force in the Rayleigh approximation calculated by Eq. (3b) in Ref. [15]. The digits in (a) indicate a multiple of the magnified forces. The triangles, stars, diamonds, circles, and hollow circles represent respectively the stable equilibrium positions of $F_e$, $F_m$, $F_{em}$, $F$, and $F_{Ray}$.

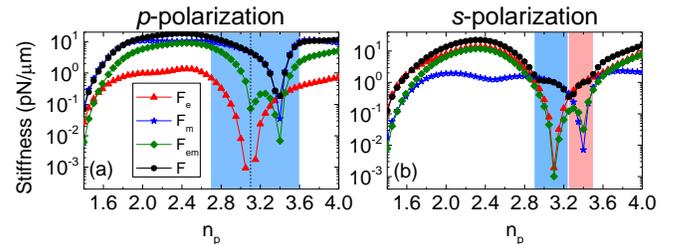

Fig. 3. The stiffness at the first stable equilibrium positions of $F$, $F_e$, $F_m$, and $F_{em}$ versus refractive index ($n_p$) of particles with $r$=150nm for $p$-polarization (a) and $s$-polarization (b). The faint blue area in (a) denotes the region where the stiffness of $F$ is dominated by that of $F_m$. The vertical black dashed line marks the location $n_p$ =3.1. The faint blue and red areas in (b) represent the regions where the stiffness of $F$ is respectively dominated by those of $F_m$ and $F_e$.

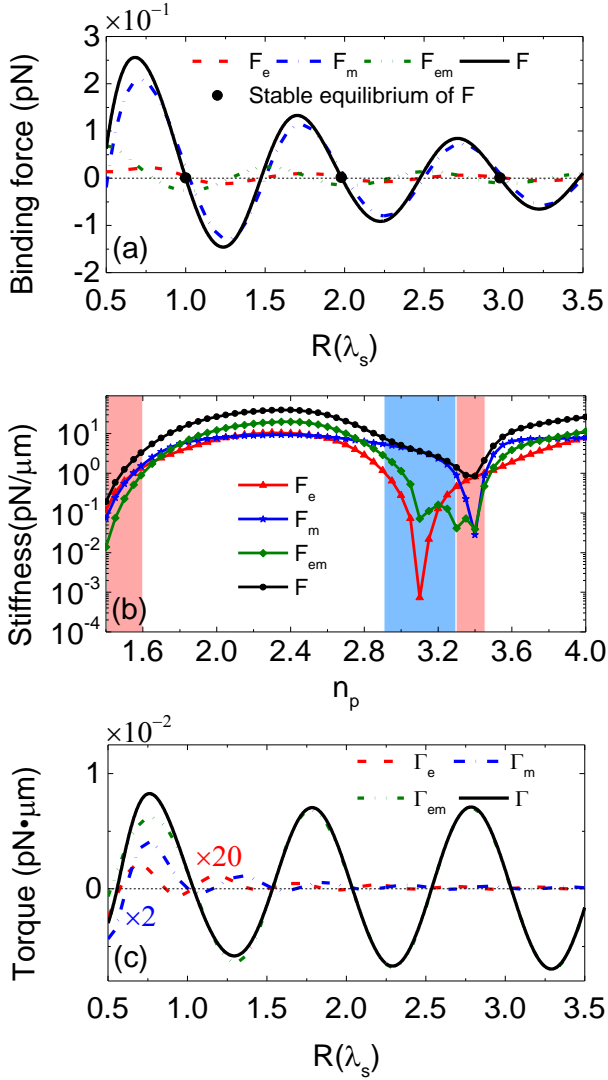

Fig. 4. Optical binding for left-circular polarization. (a) Binding force $F$ including $F_e$, $F_m$, and $F_{em}$ versus distance $R$. The black circles denote the stable equilibrium positions of $F$. (b) The stiffness at the first stable equilibrium positions of $F_e$, $F_m$, $F_{em}$, and $F$ versus $n_p$. (c) Torque $\Gamma$ including $\Gamma_e$, $\Gamma_m$, and $\Gamma_{em}$ versus distance $R$. The digits in (c) indicate a multiple of the magnified torques. The parameters in (a) and (c) are the same as in Fig. 2. The parameters in (b) are the same as in Fig. 3.

### B. The TOBF and Torque for Circularly Polarized Waves

Fig. 4 (a) shows $F$, $F_e$, $F_m$ and $F_{em}$ for left-hand circular polarization for the same parameters as in Fig. 2. The electric dipoles are again greatly suppressed compared to the magnetic dipoles and $F_m$ provides the leading contribution to $F$. As a result, the stable equilibrium positions of $F$ are very close to the counterparts of $F_m$ (not marked). This is similar to the p- and s-polarized cases in Fig. 2. Figure 4 (b) shows that the radial (along dimer's axis) stability (stiffness) of the dimer at the first radial stable equilibrium position (see details in Appendix K) is determined by different components of $F$. In the low-refractive-index region (left faint red area) the stability is determined by $F_e$. With increase of $n_p$, the stability is controlled by all components of $F$. Interestingly in the middle-refractive-index region (faint blue area) $F_m$ provides robust stability where the stiffness of $F_e$ and $F_{em}$ drops sharply. On the contrary, $F_e$ sustains the stability where the stiffness of $F_m$ and $F_{em}$ reduces sharply (right faint red area). Hence, $F_e$ and $F_m$ supplement each other in the radial stability of the dimer which is similar to the s-polarization in Fig. 3 (b).

Under circularly polarized light, a dimer experiences not only a radial force due to the TOBF, but is also subjected to a torque about the centre of mass. Figure 4 (c) shows the torque ($\Gamma$) and its components $\Gamma_e$, $\Gamma_m$, and $\Gamma_{em}$ as functions of $R$. The envelopes of $\Gamma_e$ and $\Gamma_m$ decay $\propto R^{-1}$ as shown by Eq. (11). Interestingly, the envelope of $\Gamma_{em}$ is independent of $R$ (as shown by Eq. (12)) which is similar to magnetodielectric particles [24]. Hence, $\Gamma_{em}$ plays the leading role in $\Gamma$, which is completely different to the binding force. (see details in Appendix L).

As expected, when the polarization is changed from left- to right-handed circular, the binding force stays the same but the torque in Fig. 4 (c) changes sign.

## IV. CONCLUSIONS

In summary, we have presented analytical solutions for the TOBF and torque on two identical dual dipolar dielectric particles in a plane EM wave. The electric and the magnetic dipolar interactions dominate the force in different regimes of refractive index, while the torque experienced by the dimer in circularly-polarized light is always dominated by the electric-magnetic coupling interaction. This shows that all three components must be considered for an accurate understanding of the stability and dynamics of dimers formed from high refractive index particles. Furthermore, the force, torque, and stability of the dimer with high refractive index are significantly enhanced by both magnetic and coupling interactions. The conclusions demonstrate clearly that the force, torque, and stability of the particles may be underestimated to a high degree if one uses the Rayleigh Approximation. Our results have potential applications in light-controlled self-assembly of dielectric materials of high–refractive-index. Visualizing these key differences between the different regimes of refractive index would be intriguing in an experimental optical binding geometry. In addition to the accurate determination of particle positions of the dimers there are also prospects for observing the inter-particle field directly using multiphoton excitation [41].

## ACKNOWLEDGMENTS

This work was supported by the National Key R&D Program of China (2017YFA0303400) and Postgraduate Education Reform Project of Tongji University (2018GH103). KD acknowledges support of the UK Engineering and Physical Sciences Research Council (grant EP/P030017/1).

## APPENDIX A: CALCULATION OF OPTICAL FORCE

It is assumed that a dimer consisting of two identical dielectric nanospheres is immersed in water and illuminated by an arbitrarily polarized plane EM wave. The configuration is shown by Fig. 1 in the main text. The total field encircling one particle is the sum of the incident field and the field scattered by the neighboring particle. For instance, taking into account the mutual scattering between particles, the total electric ($\mathbf{E}^B$) and magnetic ($\mathbf{H}^B$) fields surrounding particle $B$ are expressed as [30]

$$\mathbf{E}^B = \mathbf{E}_0^B + \frac{1}{\varepsilon_0 \varepsilon_s} \ddot{\mathbf{G}}_E (\mathbf{r}_B - \mathbf{r}_A) \cdot \mathbf{p}^A + iZ\ddot{\mathbf{G}}_M (\mathbf{r}_B - \mathbf{r}_A) \cdot \mathbf{m}^A, \quad (A1)$$

$$\mathbf{H}^B = \mathbf{H}_0^B - \frac{i}{Z\varepsilon_0 \varepsilon_s} \ddot{\mathbf{G}}_M (\mathbf{r}_B - \mathbf{r}_A) \cdot \mathbf{p}^A + \ddot{\mathbf{G}}_E (\mathbf{r}_B - \mathbf{r}_A) \cdot \mathbf{m}^A, \quad (A2)$$

where $\mathbf{r}_A$ and $\mathbf{r}_B$ are positions of the centers of the two particles and $Z=[\mu_0\mu_s/(\varepsilon_0\varepsilon_s)]^{1/2}$ is the impedance of the medium. $\mathbf{p}^A = \varepsilon_0\varepsilon_s\alpha_e\mathbf{E}^A$ and $\mathbf{m}^A = \alpha_m\mathbf{H}^A$ represent respectively the electric and magnetic dipole moments in particle A induced by the total electric ($\mathbf{E}^A$) and magnetic ($\mathbf{H}^A$) fields at particle A (the expressions for $\mathbf{p}$ and $\mathbf{m}$ in particle B are obtained by substituting the letter A by B). $\ddot{\mathbf{G}}_E(\mathbf{r}_B - \mathbf{r}_A)$ and $\ddot{\mathbf{G}}_M(\mathbf{r}_B - \mathbf{r}_A)$ are free-space electric and magnetic dyadic Green's functions of a point dipole [42] (see details in Appendix B). The first term in Eq. (A1) [Eq. (A2)] represents the incident electric (magnetic) field at particle B, the second one denotes the scattered electric (magnetic) field on particle B by the electric dipole in particle A, and the last term expresses the scattered electric (magnetic) field on particle B by the magnetic dipole in particle A. Additionally, the total electric and magnetic fields in the nearby particle A are also described by Eqs. (A1) and (A2) through interchanging the superscripts B and A while considering the relations $\ddot{\mathbf{G}}_E(\mathbf{r}_A - \mathbf{r}_B) = \ddot{\mathbf{G}}_E(\mathbf{r}_B - \mathbf{r}_A)$ and $\ddot{\mathbf{G}}_M(\mathbf{r}_A - \mathbf{r}_B) = -\ddot{\mathbf{G}}_M(\mathbf{r}_B - \mathbf{r}_A)$. Hereafter we simplify the notation $\ddot{\mathbf{G}}_E(\mathbf{r}_B - \mathbf{r}_A) \equiv \ddot{\mathbf{G}}_E$ and $\ddot{\mathbf{G}}_M(\mathbf{r}_B - \mathbf{r}_A) \equiv \ddot{\mathbf{G}}_M$ for convenience. By decomposing vector Eqs. (A1) and (A2) into scalar equations along the three axes and solving them for particles A and B in terms of $\mu$, $\kappa$, and $\eta$, the $x$, $y$, and $z$ components of the electric and magnetic dipole moments in each particle are derived in Eqs. (A7)-(A12).

It is convenient to define dressed polarizabilities to describe the polarization of the particles under the incoming waves. In detail, the electric dressed polarizabilities are described by

$$\tilde{\alpha}_{ex} = \frac{\alpha_e}{1-\alpha_e\mu}, \quad \tilde{\alpha}_{ey} = \frac{\alpha_e}{1-\alpha_e\kappa}, \quad \tilde{\alpha}_{ez} = \frac{\alpha_e}{1+\alpha_e\mu}, \quad (A3)$$

where $\tilde{\alpha}_{ev}$ ($v=x, y,$ and $z$) is caused by the hybridization of two $v$ components of the electric dipole moments ($p_v$) in different particles. On the other hand, the magnetic dressed polarizibilities are expressed as

$$\tilde{\alpha}_{mx} = \frac{\alpha_m}{1-\alpha_m\mu}, \quad \tilde{\alpha}_{my} = \frac{\alpha_m}{1-\alpha_m\kappa}, \quad \tilde{\alpha}_{mz} = \frac{\alpha_m}{1+\alpha_m\mu}, \quad (A4)$$

where $\tilde{\alpha}_{mv}$ is the result of hybridization between $v$ components of the magnetic dipole moments ($m_v$) in two particles. In addition, the $p$-polarized dressed polarizabilities are given by

$$\tilde{\alpha}_{em-p,1} = \frac{\tilde{\alpha}_{ez}\tilde{\alpha}_{mx}\eta}{1-\tilde{\alpha}_{ez}\tilde{\alpha}_{mx}\eta^2}, \quad \tilde{\alpha}_{em-p,2} = \frac{\tilde{\alpha}_{mx}}{1-\tilde{\alpha}_{ez}\tilde{\alpha}_{mx}\eta^2}, \quad (A5)$$

where $\tilde{\alpha}_{em-p,1}$ originates from the hybridization of $p_z$ in one particle with $p_z$ and $m_x$ in another while $\tilde{\alpha}_{em-p,2}$ is a result of the hybridization of $m_x$ in one particle with $m_x$ and $p_z$ in other one. They can be thought of being induced by the $p$-polarized wave ($\mathbf{H}_0$ is perpendicular to the incident plane and along $x$ axis) which causes the $m_x$ in the two particles. Moreover, the $s$-polarized dressed polarizabilities are written as

$$\tilde{\alpha}_{em-s,1} = \frac{\tilde{\alpha}_{ex}\tilde{\alpha}_{mz}\eta}{1-\tilde{\alpha}_{ex}\tilde{\alpha}_{mz}\eta^2}, \quad \tilde{\alpha}_{em-s,2} = \frac{\tilde{\alpha}_{ex}}{1-\tilde{\alpha}_{ex}\tilde{\alpha}_{mz}\eta^2}, \quad (A6)$$

where $\tilde{\alpha}_{em-s,1}$ comes from the hybridization of $m_z$ in one particle with $m_z$ and $p_x$ in the other, while $\tilde{\alpha}_{em-s,2}$ stems from the hybridization of $p_x$ in one particle with $p_x$ and $m_z$ in the neighboring one. They can be regarded as a result of the $s$-polarized wave ($\mathbf{E}_0$ is perpendicular to the incident plane and along $x$ axis) which induces the $p_x$ in the two particles. The self-consistent electric and magnetic dipole moments in the two particles induced by an arbitrarily polarized EM wave and hybridization between the particles are expressed as

$$p_x^A = p_x^B = \varepsilon_0\varepsilon_s\tilde{\alpha}_{em-s,2}E_{0,x}, \quad (A7)$$

$$p_y^A = p_y^B = \varepsilon_0\varepsilon_s\tilde{\alpha}_{ey}E_{0,y}, \quad (A8)$$

$$p_z^A = -p_z^B = i\varepsilon_0\varepsilon_s\tilde{\alpha}_{em-p,1}E_{0,y}, \quad (A9)$$

$$m_x^A = m_x^B = \tilde{\alpha}_{em-p,2}H_{0,x}, \quad (A10)$$

$$m_y^A = m_y^B = \tilde{\alpha}_{my}H_{0,y}, \quad (A11)$$

$$m_z^A = -m_z^B = i\tilde{\alpha}_{em-s,1}H_{0,y}. \quad (A12)$$

Equation (A7) denotes that $p_x$ in one particle is partly excited by the incident electric field $E_{0,x}$ and partly hybridized by both $p_x$ and $m_z$ in neighboring particle. Equation (A8) indicates that $p_y$ in one particle is hybridized by $p_y$ in the other in addition to the induction by the incident electric field $E_{0,y}$. Equation (A9) shows that $p_z$ in one particle is hybridized by both $p_z$ and $m_x$ in the other particle, rather than by the induction of the incident electric field. Equation (A10) reveals that $m_x$ in one particle is not only induced by the incident magnetic field $H_{0,x}$ but also hybridized by both $p_z$ and $m_x$ in the other particle. Equation (A11) exhibits that $m_y$ in one particle is caused by the incident magnetic field $H_{0,y}$ and hybridized by $m_y$ in the neighboring particle. Equation (A12) denotes that $m_z$ in one particle is hybridized by both $p_x$ and $m_z$ in the neighboring particle and independent of the incident magnetic field (see details in Appendix C). For a $p$-polarized wave, $\tilde{\alpha}_{my}$, $\tilde{\alpha}_{mz}$, $\tilde{\alpha}_{ex}$, $\tilde{\alpha}_{em-s,1}$, and $\tilde{\alpha}_{em-s,2}$ disappear as well as $p_x^{A(B)} = m_y^{A(B)} = m_z^{A(B)} = 0$ due to the absence of $E_{0,x}$ and $H_{0,y}$. On the other hand, for s-polarization without $E_{0,y}$ and $H_{0,x}$, the $\tilde{\alpha}_{mx}$, $\tilde{\alpha}_{ey}$, $\tilde{\alpha}_{ez}$, $\tilde{\alpha}_{em-p,1}$, and $\tilde{\alpha}_{em-p,2}$ vanish while $p_y^{A(B)} = p_z^{A(B)} = m_x^{A(B)} = 0$.

For calculation of the optical force, taking particle B as an example, the spatial derivatives of the total electric and magnetic fields at particle B are be solved as

$$\frac{\partial \mathbf{E}^B}{\partial u} = \frac{\partial \mathbf{E}_0^B}{\partial u} + \frac{1}{\varepsilon_0 \varepsilon_s} \frac{\partial \ddot{\mathbf{G}}_E}{\partial u} \cdot \mathbf{p}^A + iZ \frac{\partial \ddot{\mathbf{G}}_M}{\partial u} \cdot \mathbf{m}^A, \quad (A13)$$

$$\frac{\partial \mathbf{H}^B}{\partial u} = \frac{\partial \mathbf{H}_0^B}{\partial u} - \frac{i}{Z\varepsilon_0 \varepsilon_s} \frac{\partial \ddot{\mathbf{G}}_M}{\partial u} \cdot \mathbf{p}^A + \frac{\partial \ddot{\mathbf{G}}_E}{\partial u} \cdot \mathbf{m}^A, \quad (A14)$$

where $\partial/\partial u$ denotes the partial derivative with respect to the arguments $x$ and $y$. The partial derivatives of the located fields $\mathbf{E}$ and $\mathbf{H}$ at particle A also can be calculated by Eqs. (A13) and (A14) through exchanging the letter A and B while considering the relations $\partial \ddot{\mathbf{G}}_E/\partial u\big|_A = -\partial \ddot{\mathbf{G}}_E/\partial u\big|_B$ and $\partial \ddot{\mathbf{G}}_M/\partial u\big|_A = \partial \ddot{\mathbf{G}}_M/\partial u\big|_B$. The matrix forms of $\ddot{\mathbf{G}}_{E(M)}$ and $\partial \ddot{\mathbf{G}}_{E(M)}/\partial u$, and the scalar decompositions of Eqs. (A13) and (A14) are presented in Appendixes D and E. The time-averaged optical force on a dielectric particle with induced electric and magnetic dipoles is given by [29]

$$\mathbf{F} = \frac{1}{2} \Re \left[ (\mathbf{p}\nabla) \mathbf{E}^* + \mu_0 \mu_s (\mathbf{m}\nabla) \mathbf{H}^* - \frac{Zk^4}{6\pi} (\mathbf{p} \times \mathbf{m}^*) \right]. \quad (A15)$$

The first two terms on the right-hand of Eq. (A15) are respectively the forces on the electric and magnetic dipoles while the last one is the electric-magnetic coupling force due to the interference between the two dipoles.

## APPENDIX B: EIGENVALUES OF $\ddot{\mathbf{G}}_E$ AND $\ddot{\mathbf{G}}_M$.

The free-space electric ($\ddot{\mathbf{G}}_E$) and magnetic ($\ddot{\mathbf{G}}_M$) dyadic Green's functions are written as [42]

$$\ddot{\mathbf{G}}_E = \frac{\exp(ikR)}{4\pi R^3} \left[ (3 - 3ikR - k^2 R^2) \frac{\mathbf{RR}}{R^2} + (-1 + ikR + k^2 R^2) \ddot{\mathbf{I}} \right], \quad (B1)$$

$$\ddot{\mathbf{G}}_M = \frac{\exp(ikR)}{4\pi R^3} (ik^2 R^2 - kR) \frac{\mathbf{R} \times \ddot{\mathbf{I}}}{R}, \quad (B2)$$

where $\ddot{\mathbf{I}}$ is unit dyad. Eqation (B2) is derived from Eq. (B1) by utilizing the orthogonality relation $\ddot{\mathbf{G}}_M = (\nabla \times \ddot{\mathbf{G}}_E)/k$. For simplified calculations of the induced dipole moments in particles and TOBF, $\ddot{\mathbf{G}}_E$ is expressed in terms of the eigenequation [22] as

$$\ddot{\mathbf{G}}_E = (\kappa - \mu) \frac{\mathbf{RR}}{R^2} + \mu \ddot{\mathbf{I}}, \quad (B3)$$

where the eigenvalues $\kappa$ and $\mu$ of $\ddot{\mathbf{G}}_E$ are determined by

$$\mu = \frac{\exp(ikR)}{4\pi R^3} (k^2 R^2 + ikR - 1), \quad (B4)$$

$$\kappa = \frac{\exp(ikR)}{4\pi R^3} (-2ikR + 2). \quad (B5)$$

Since any vector $\mathbf{V}$ can be decomposed into two components $\mathbf{V}_{\parallel}$ and $\mathbf{V}_{\perp}$ who are respectively parallel and perpendicular to vector $\mathbf{R}$, based on Eq. (B3), the dot product between $\ddot{\mathbf{G}}_E$ and any vector $\mathbf{V}$ is given by

$$\ddot{\mathbf{G}}_E \cdot \mathbf{V} = \kappa \mathbf{V}_{//} + \mu \mathbf{V}_{\perp}, \quad (B6)$$

which denotes that the eigenvectors of $\kappa$ and $\mu$ are respectively parallel and orthogonal to vector $\mathbf{R}$. Then, we have

$$\ddot{\mathbf{G}}_E \cdot (\mathbf{e}_x, \mathbf{e}_y, \mathbf{e}_z) = (\mu \mathbf{e}_x, \kappa \mathbf{e}_y, \mu \mathbf{e}_z), \quad (B7)$$

where $\mathbf{e}_x$, $\mathbf{e}_y$, and $\mathbf{e}_z$ are unit vectors along $x$, $y$, and $z$ axes in the Cartesian system. On the other hand, the eigenequation of $\ddot{\mathbf{G}}_M$ is written as

$$\ddot{\mathbf{G}}_M = \eta \mathbf{R} \times \ddot{\mathbf{I}}/R, \quad (B8)$$

where the eigenvalue $\eta$ of $\ddot{\mathbf{G}}_M$ is expressed as

$$\eta = \frac{\exp(ikR)}{4\pi R^3} (ik^2 R^2 - kR). \quad (B9)$$

Equation (B8) demonstrates that the eigenvector of $\eta$ is orthogonal to vector $\mathbf{R}$. Analogous to Eq. (B6), based on Eq. (B8), the dot product between $\ddot{\mathbf{G}}_M$ and any vector $\mathbf{V}$ is expressed by

$$\ddot{\mathbf{G}}_M \cdot \mathbf{V} = \eta \mathbf{V}_{\perp}. \quad (B10)$$

Hence, the dot product between $\ddot{\mathbf{G}}_M$ and unit vectors is expressed as

$$\ddot{\mathbf{G}}_M \cdot (\mathbf{e}_x, \mathbf{e}_y, \mathbf{e}_z) = (-\eta \mathbf{e}_z, 0, \eta \mathbf{e}_x). \quad (B11)$$

## APPENDIX C: INDUCED DIPOLE MOMENTS IN DIMER

By substituting Eqs. (B7) and (B11) into Eq. (A1), the three components of the total electric field on particle A (B) are expressed by

$$E_x^{A(B)} = E_{0,x}^{A(B)} + \frac{1}{\varepsilon_0 \varepsilon_s} \mu p_x^{B(A)} \mp iZ\eta m_z^{B(A)}, \quad (C1)$$

$$E_y^{A(B)} = E_{0,y}^{A(B)} + \frac{1}{\varepsilon_0 \varepsilon_s} \kappa p_y^{B(A)}, \quad (C2)$$

$$E_z^{A(B)} = \frac{1}{\varepsilon_0 \varepsilon_s} \mu p_z^{B(A)} \pm iZ\eta m_x^{B(A)}. \quad (C3)$$

Notice that for electric field on particle A, the signs of the third term in Eq. (C1) and the second term in Eq. (C3) are respectively negative and positive while they are opposite for particle B. The regulations of the signs are also appropriate for the following Eqs. (C4) and (C6). Considering the relation $\mathbf{p}=\varepsilon_0\varepsilon_s\alpha_e\mathbf{E}$, the three components of the electric dipole moment in particle A (B) are given by

$$p_x^{A(B)} = \varepsilon_0\varepsilon_m\alpha_e E_{0,x}^{A(B)} + \alpha_e\mu p_x^{B(A)} \mp iZ\varepsilon_0\varepsilon_m\alpha_e\eta m_z^{B(A)}, \quad (C4)$$

$$p_y^{A(B)} = \varepsilon_0\varepsilon_m\alpha_e E_{0,y}^{A(B)} + \alpha_e\kappa p_y^{B(A)}, \quad (C5)$$

$$p_z^{A(B)} = \alpha_e\mu p_z^{B(A)} \pm iZ\varepsilon_0\varepsilon_m\alpha_e\eta m_x^{B(A)}, \quad (C6)$$

where $p_\nu^{A(B)}$ and $m_\nu^{A(B)}$ ($\nu = x$, $y$, and $z$) represent respectively the three components of the electric and magnetic dipole moments induced in particle A (B); $E_{0,\nu}$ and $H_{0,\nu}$ stand for the incident electric and magnetic fields along the three axes. In detail, Eq. (C4) shows that $p_x$ in one particle is excited by the incident electric field $E_{0,x}$ (first term) and the re-radiated electric fields by both $p_x$ (second term) and $m_z$ (the third terms) in the neighboring particle. Equation (C5) denotes that $p_y$ in one particle is caused by the incident electric field $E_{0,y}$ (first term) and the re-excited electric field by $p_y$ (second term) in the other. Equation (C6) illuminates that $p_z$ in one particle originates from the re-radiated electric fields by both $p_z$ (first term) and $m_x$ (second term) in the other particle.

On the other hand, substituting Eqs. (B7) and (B11) into Eq. (A2), the three components of the total magnetic field on particle A (B) are expressed by

$$H_x^{A(B)} = H_{0,x}^{A(B)} \pm \frac{i}{Z\varepsilon_0\varepsilon_s}\eta p_z^{B(A)} + \mu m_x^{B(A)}, \quad (C7)$$

$$H_y^{A(B)} = H_{0,y}^{A(B)} + \kappa m_y^{B(A)}, \quad (C8)$$

$$H_z^{A(B)} = \mp\frac{i}{Z\varepsilon_0\varepsilon_s}\eta p_x^{B(A)} + \mu m_z^{B(A)}. \quad (C9)$$

Notice that for magnetic field on particle A, the signs of the second term in Eq. (C7) and the first term in Eq. (C9) are respectively positive and negative while they are opposite for particle B. The regulations of the signs are also appropriate for the following Eqs. (C10) and (C12). Owing to the relation $\mathbf{m}=\alpha_m\mathbf{H}$, the three components of the magnetic dipole moment in particle A (B) are determined by

$$m_x^{A(B)} = \alpha_m H_{0,x}^{A(B)} \pm \frac{i}{Z\varepsilon_0\varepsilon_m}\alpha_m\eta p_z^{B(A)} + \alpha_m\mu m_x^{B(A)}, \quad (C10)$$

$$m_y^{A(B)} = \alpha_m H_{0,y}^{A(B)} + \alpha_m\kappa m_y^{B(A)}, \quad (C11)$$

$$m_z^{A(B)} = \mp\frac{i}{Z\varepsilon_0\varepsilon_m}\alpha_m\eta p_x^{B(A)} + \alpha_m\mu m_z^{B(A)}. \quad (C12)$$

Equation (C10) shows that $m_x$ in one particle is induced by the incident magnetic field $H_{0,x}$ (first term) and the re-excited magnetic fields by both $p_z$ (second term) and $m_x$ (last term) in the other. Equation (C11) indicates that the $m_y$ in one particle is caused by the incident magnetic field $H_{0,y}$ (first term) and the re-radiated magnetic field by $m_y$ (second term) in the neighboring particle. Equation (C12) indicated that $m_z$ in one particle comes from the re-excited magnetic fields by $p_x$ (first term) and $m_z$ (second term) in the neighboring particle. Notice that the $p_z$ and $m_z$ are not directly excited by the incident electric and magnetic fields without a non-zero $z$ component. Equations (C4)-(C6) and (C10)-(C12) demonstrate clearly the hybridizations between the electric and magnetic dipoles in different particles.

**APPENDIX D: MATRIX AND SPATIAL PARTIAL DERIVATIVES OF $\ddot{\mathbf{G}}_E$ AND $\ddot{\mathbf{G}}_M$.**

For calculating the TOBF, the matrix forms and spatial partial derivatives of $\ddot{\mathbf{G}}_E$ and $\ddot{\mathbf{G}}_M$ must be known. Substituting the tensors $\mathbf{RR}/R^2$ and $\ddot{\mathbf{I}}$ into Eq. (B3), the matrix of $\ddot{\mathbf{G}}_E$ in terms of $\kappa$ and $\mu$ is expressed as

$$\ddot{\mathbf{G}}_E = \begin{pmatrix} \mu & 0 & 0 \\ 0 & \kappa & 0 \\ 0 & 0 & \mu \end{pmatrix}. \quad (D1)$$

In addition, based on Eq. (B3) and (D1), the spatial partial derivatives of $\ddot{\mathbf{G}}_E$ with respect to the $x$ and $y$ axes are given by

$$\frac{\partial\ddot{\mathbf{G}}_E}{\partial x} = \frac{\kappa-\mu}{R}\begin{pmatrix} 0 & 1 & 0 \\ 1 & 0 & 0 \\ 0 & 0 & 0 \end{pmatrix}, \quad (D2)$$

$$\frac{\partial\ddot{\mathbf{G}}_E}{\partial y} = \begin{pmatrix} \partial\mu/\partial R & 0 & 0 \\ 0 & \partial\kappa/\partial R & 0 \\ 0 & 0 & \partial\mu/\partial R \end{pmatrix}. \quad (D3)$$

On the other hand, we have

$$\left(\frac{\mathbf{R}\times\ddot{\mathbf{I}}}{R}\right) = \begin{pmatrix} 0 & 0 & -1 \\ 0 & 0 & 0 \\ 1 & 0 & 0 \end{pmatrix}, \quad (D4)$$

where $\mathbf{R}\times\ddot{\mathbf{I}}$ denotes the matrix generated by the cross-product of $\mathbf{R}$ with each column vector of $\ddot{\mathbf{I}}$ [42]. Substituting Eq. (D4) into Eq. (B8), the matrix of $\ddot{\mathbf{G}}_M$ in terms of $\eta$ is described as

$$\ddot{\mathbf{G}}_M = \eta \begin{pmatrix} 0 & 0 & -1 \\ 0 & 0 & 0 \\ 1 & 0 & 0 \end{pmatrix}. \quad (D5)$$

Moreover, based on Eqs. (B8) and (D5), the spatial partial derivative of $\ddot{\mathbf{G}}_M$ with respect to the $x$ and $y$ axes are read as

$$\frac{\partial \ddot{\mathbf{G}}_M}{\partial x} = \frac{\eta}{R} \begin{pmatrix} 0 & 0 & 0 \\ 0 & 0 & 1 \\ 0 & -1 & 0 \end{pmatrix}, \quad (D6)$$

$$\frac{\partial \ddot{\mathbf{G}}_M}{\partial y} = \frac{\partial \eta}{\partial R} \begin{pmatrix} 0 & 0 & -1 \\ 0 & 0 & 0 \\ 1 & 0 & 0 \end{pmatrix}. \quad (D7)$$

Notice that $\partial \vartheta / \partial x = \partial \vartheta / \partial z = 0$ and $\partial \vartheta / \partial y = \partial \vartheta / \partial R$ ($\vartheta = \kappa$, $\mu$, and $\eta$) because $R$ is along the $y$ axis.

### APPENDIX E: SPATIAL PARTIAL DERIVATIVE OF TOTAL $E^B$ AND $H^B$

Let's take particle B as an example. First, Eq. (A13) is decomposed into scalar equations along the three Cartesian axes. Second, Eqs. (D2), (D3), (D6), and (D7) are substituted into the three components of Eq. (A13) while using the relation $\partial \xi_{0,u}^B / \partial x = \partial \xi_{0,u}^B / \partial y = 0$ ($\zeta = E$ and $H$, $u = x$ and $y$) for a plane wave propagating along the $z$ axis. Finally, the spatial partial derivative of the total electric field in the nearby particle B with respect to the $x$ axis are expressed as

$$\frac{\partial E_x^B}{\partial x} = \frac{\kappa - \mu}{\varepsilon_0 \varepsilon_s R} p_y^A, \quad (E1)$$

$$\frac{\partial E_y^B}{\partial x} = \frac{\kappa - \mu}{\varepsilon_0 \varepsilon_s R} p_x^A + iZ \frac{\eta}{R} m_z^A, \quad (E2)$$

$$\frac{\partial E_z^B}{\partial x} = -iZ \frac{\eta}{R} m_y^A. \quad (E3)$$

In addition, the spatial partial derivative of the total electric field along the $y$ axis are written as

$$\frac{\partial E_x^B}{\partial y} = \frac{1}{\varepsilon_0 \varepsilon_s} \frac{\partial \mu}{\partial y} p_x^A - iZ \frac{\partial \eta}{\partial y} m_z^A, \quad (E4)$$

$$\frac{\partial E_y^B}{\partial y} = \frac{1}{\varepsilon_0 \varepsilon_s} \frac{\partial \kappa}{\partial y} p_y^A, \quad (E5)$$

$$\frac{\partial E_z^B}{\partial y} = \frac{1}{\varepsilon_0 \varepsilon_s} \frac{\partial \mu}{\partial y} p_z^A + iZ \frac{\partial \eta}{\partial y} m_x^A. \quad (E6)$$

Analogously, Eq. (A14) is decomposed into scalar equations along the three axes while Eqs. (D2), (D3), (D6), and (D7) are substituted into the three components of Eq. (A14). The spatial partial derivative of the total magnetic field in the nearby particle B with respect to the $x$ axis are expressed by

$$\frac{\partial H_x^B}{\partial x} = \frac{\kappa - \mu}{R} m_y^A, \quad (E7)$$

$$\frac{\partial H_y^B}{\partial x} = -\frac{i}{Z \varepsilon_0 \varepsilon_s} \frac{\eta}{R} p_z^A + \frac{\kappa - \mu}{R} m_y^A, \quad (E8)$$

$$\frac{\partial H_z^B}{\partial x} = \frac{i}{Z \varepsilon_0 \varepsilon_s} \frac{\eta}{R} p_y^A. \quad (E9)$$

Meanwhile, the spatial partial derivative of the total magnetic field along the $y$ axis are described as

$$\frac{\partial H_x^B}{\partial y} = \frac{i}{Z \varepsilon_0 \varepsilon_s} \frac{\partial \eta}{\partial y} p_z^A + \frac{\partial \mu}{\partial y} m_x^A, \quad (E10)$$

$$\frac{\partial H_y^B}{\partial y} = \frac{\partial \kappa}{\partial y} m_y^A, \quad (E11)$$

$$\frac{\partial H_z^B}{\partial y} = -\frac{i}{Z \varepsilon_0 \varepsilon_s} \frac{\partial \eta}{\partial y} p_x^A + \frac{\partial \mu}{\partial y} m_z^A. \quad (E12)$$

### APPENDIX F: TOBF AND TORQUE

By substituting Eqs. (A7)-(A12) and Eqs. (E1)-(E12) into Eq. (A15), the electric component of the binding force along the dimer axis is expressed as

$$F_e = \frac{\varepsilon_0 \varepsilon_s}{2} \left\{ \begin{array}{l} \Re\left[\frac{\partial \kappa}{\partial R}\right] |\tilde{\alpha}_{ey}|^2 |E_{0,y}|^2 - \Re\left[\frac{\partial \mu}{\partial R}\right] |\tilde{\alpha}_{em-p,1}|^2 |E_{0,y}|^2 \\ + \Re\left[\frac{\partial \eta}{\partial R} \tilde{\alpha}_{em-p,2} \tilde{\alpha}_{em-p,1}^*\right] |E_{0,y}|^2 \\ + \Re\left[\frac{\partial \eta}{\partial R} \tilde{\alpha}_{em-s,1} \tilde{\alpha}_{em-s,2}^*\right] |E_{0,x}|^2 + \Re\left[\frac{\partial \mu}{\partial R}\right] |\tilde{\alpha}_{em-s,2}|^2 |E_{0,x}|^2 \end{array} \right\},$$

(F1)

the magnetic component of the binding force is expressed as

$$F_m = \frac{\varepsilon_0 \varepsilon_s}{2} \left\{ \begin{array}{l} +\Re\left[\frac{\partial \mu}{\partial R}\right] |\tilde{\alpha}_{em-p,2}|^2 |E_{0,y}|^2 + \Re\left[\frac{\partial \eta}{\partial R} \tilde{\alpha}_{em-p,1} \tilde{\alpha}_{em-p,2}^*\right] |E_{0,y}|^2 \\ \Re\left[\frac{\partial \kappa}{\partial R}\right] |\tilde{\alpha}_{my}|^2 |E_{0,x}|^2 - \Re\left[\frac{\partial \mu}{\partial R}\right] |\tilde{\alpha}_{em-s,1}|^2 |E_{0,x}|^2 \\ +\Re\left[\frac{\partial \eta}{\partial R} \tilde{\alpha}_{em-s,2}^2 \tilde{\alpha}_{em-s,1}^*\right] |E_{0,x}|^2 \end{array} \right\},$$

(F2)

and the electric-magnetic coupling component of the binding force is given by

$$F_{em} = \frac{\varepsilon_0 \varepsilon_s}{12\pi} k^4 \Im\left[ \tilde{\alpha}_{em-p,1} \tilde{\alpha}^*_{em-p,2} |E_{0,y}|^2 - \tilde{\alpha}_{em-s,2} \tilde{\alpha}^*_{em-s,1} |E_{0,x}|^2 \right]. \quad (F3)$$

On the other hand, the electric component of the torque along the z axis is expressed as

$$\Gamma_e = \frac{\varepsilon_0 \varepsilon_s}{2} \left\{ \begin{array}{l} 2\Re[\kappa - \mu] \Re\left[ \tilde{\alpha}_{em-s,2} \tilde{\alpha}^*_{ey} E_{0,x} E^*_{0,y} \right] \\ + \Re\left[ \eta \left( \tilde{\alpha}_{my} \tilde{\alpha}^*_{em-p,1} - \tilde{\alpha}_{em-s,2} \tilde{\alpha}^*_{ey} \right) E_{0,x} E^*_{0,y} \right] \end{array} \right\}, \quad (F4)$$

the magnetic component of the torque is given by

$$\Gamma_m = \frac{\varepsilon_0 \varepsilon_s}{2} \left\{ \begin{array}{l} -2\Re[\kappa - \mu] \Re\left[ \tilde{\alpha}_{em-p,2} \tilde{\alpha}^*_{my} E_{0,y} E^*_{0,x} \right] \\ + \Re\left[ \eta \left( \tilde{\alpha}_{em-p,1} \tilde{\alpha}^*_{my} - \tilde{\alpha}_{ey} \tilde{\alpha}^*_{em-s,1} \right) E_{0,y} E^*_{0,x} \right] \end{array} \right\}, \quad (F5)$$

and the electric-magnetic coupling component of the torque is expressed as

$$\Gamma_{em} = \frac{\varepsilon_0 \varepsilon_s}{12\pi} k^4 \Im\left[ \left( \tilde{\alpha}_{ey} \tilde{\alpha}^*_{em-s,1} + \tilde{\alpha}_{em-p,1} \tilde{\alpha}^*_{my} \right) E_{0,y} E^*_{0,x} \right]. \quad (F6)$$

## APPENDIX G: BINDING FORCE FOR LOW- AND MODERATE-INDEX PARTICLES.

Figure S1 (a) shows that $F_m$ (blue solid curve) and $F_{em}$ (green solid curve) have non-ignorable contributions to $F$ (black solid curve) compared to $F_e$ (red solid curve) even for a low-refractive-index dielectric dimer ($n_p$=1.59), i.e., polystyrene spheres [43] for p-polarized wave. In detail, although $F$ is mainly determined by $F_e$ when $R$ is smaller than the first stable equilibrium position of $F$ (the first solid black circle on $R$ axis), the contributions of $F_m$ and $F_{em}$ to $F$ cannot be completely ignored. In particular, the first stable equilibrium position of $F$ coincides with the counterpart of $F_{em}$ (the first solid green diamond) instead of that of $F_e$ (the first solid red triangle) even though $F_e$ is an order of magnitude larger than $F_{em}$. In addition, as $R$ increases, the decay of $F_{em}$ is slower than that of $F_m$ while the decay of $F_m$ is slower than that of $F_e$. The reasons are the same as in Fig. 2 (a) and as shown in Eqs. (7)-(9) in the main text. The phenomena also exist in Figs. S1 (b). Consequently, the contribution of $F_m$ to $F$ is as much as that of $F_e$, when $R$ goes beyond the first stable equilibrium position of $F_e$. As a result, the second stable equilibrium position of $F$ (the second solid black circle) diverges from the counterpart of $F_{em}$ (the second solid green diamond) and approaches to the corresponding equilibrium position of $F_m$ (the second solid blue star). This phenomenon is more obvious in the vicinity of the third stable equilibrium position of $F$ (the third solid black point). Additionally, $F_{Ray}$ overlaps completely $F_e$, while the stable equilibrium positions of $F_{Ray}$ (hollow purple circles) are consistent with those of $F_e$. The reason is that, for a low-refractive-index dielectric dimer, the hybridization between the two particles is much weaker. It can be seen that $F_{em}$ is an order of magnitude smaller than $F_e$ and $F_m$. Generally, the x component of magnetic dipole moment $m_x$ in one particle excites the z component of the electric dipole moment $p_z$ in the other due to the hybridization. But this hardly occurs for low-refractive-index particles because of the weak magnetic response. Hence, $F_e$ originates just from the electric dipole-dipole interaction between two particles while each electric dipole is induced by the incident electric field and re-excited electric field of the neighboring electric dipole. The physical mechanism is identical to $F_{Ray}$.

Interestingly, for particles with moderate refractive index $n_p$=3.5, i.e., silicon spheres [40], $F_m$ and $F_{em}$ are enhanced to exceed $F_e$ as shown in Fig. S1 (b). As expected, the stable equilibrium positions of $F$ are determined by all components of $F$, and therefore the positions deviate largely from the corresponding stable equilibrium positions of $F_e$. Importantly, $F$ has been dramatically strengthened by one order of magnitude compared to the particles with $n_p$=1.59 in Fig. S1 (a). As a result, the stability of the dimer at the stable equilibrium positions is improved by one order of magnitude. The reason is the enhanced magnetic response and hybridization of the dimer with increase of the refractive index of particles. It is worth noting that $F_{Ray}$ overestimates (underestimates) $F_e$ at the first dip (peak) of $F_e$ in front of the first stable equilibrium position of $F$ (see details in Appendix J) Additionally, the stable equilibrium positions of $F_e$ move to large $R$ compared to that of $F_{Ray}$ (hollow purple circles).

Fig. S1(c) and (d) show the TOBF and its three components for s-polarized wave. For particles with $n_p$=1.59 in Fig. 2(c), $F_m$ and $F_{em}$ are an order of magnitude smaller than $F_e$. Therefore, $F$ is almost completely determined by $F_e$ and the stable equilibrium positions of $F$ and $F_e$ are in agreement. Physically, for an s-polarized wave, the electric dipoles induced in the two particles are side-by-side parallel and directed along the x axis. On the other hand, the y components of the two magnetic dipole induced by both the incident magnetic field and hybridization between them are head-to-tail colinear. As we know, the radiated field by a dipole focuses mainly on the direction perpendicular to the dipole moment. Hence, the radiative interaction between two $p_x$ is stronger than that between two $m_y$. Additionally, $m_z$ in each particle is caused by the induced $p_x$ in the neighboring one. However, the secondary interaction between two $m_z$ is much smaller than not only the interaction between $p_x^A$ and $p_x^B$ but also the interaction between $m_y^A$ and $m_y^B$. Furthermore, every corresponding stable equilibrium position of $F_e$ and $F_{Ray}$ overlaps. Meanwhile, the $F_e$ matches completely with $F_{Ray}$ when $R$ exceeds $\lambda_s$. The reason is that the two forces originate from the same electric dipolar interaction between $p_x^A$ and $p_x^B$. What is different, however, to the case of p-polarization is that the decay of $F_{em}$ is the slowest while that of $F_m$ is the fastest with increase of $R$. The reasons are that $F_m$ is proportional to $R^{-2}$ while $F_e$ and $F_{em}$ are proportional to $R^{-1}$ in the far-field region. The phenomena also exist for moderate-refractive-index particles in Fig. S1(d).

With an increase of the refractive index of particles to $n_p$=3.5 as in Fig. S1 (d), $F$ and $F_{Ray}$ are dramatically enhanced by about 50 times compared to Fig. S1 (c). As a result, the stability of the dimer at the stable equilibrium positions of $F$ are greatly enhanced, the same as the p-polarization in Fig. S1 (b). In addition, $F_{em}$ is heightened to half of $F_e$ compared to the particle with $n_p$=1.59 in Fig. S1 (c) due to the enhanced hybridization. But $F_e$ is still an order of magnitude larger than $F_m$ because of the unchanged configurations of electric and magnetic dipoles in the dimer. As a result, $F$ is almost determined by both $F_e$ and $F_{em}$ while the stable equilibrium positions of $F$ are nearly dominated by the counterparts of $F_e$. Notice that $F_{Ray}$ underestimates (overestimates) $F_e$ at the first dip (peak) of $F_e$ although the two forces are coincident at large $R$, which is contrary to the $p$-polarization in Fig. S1 (c), (see details in Appendix H).

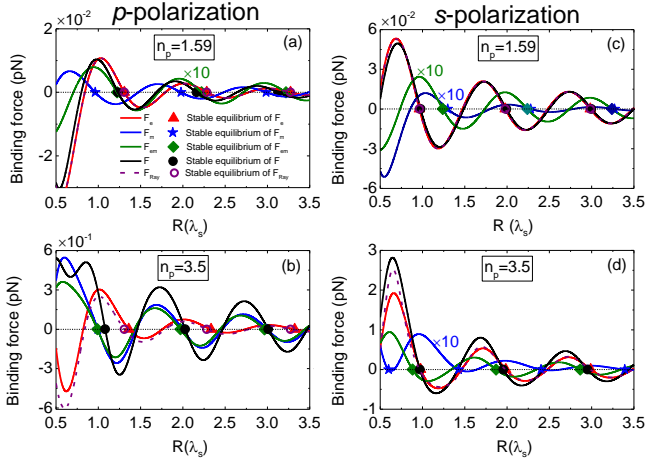

Fig. S1. Binding force upon sphere B with radius 100nm versus distance ($R$) between the centers of two spheres with refractive index $n_p$=1.59 (a), (c) and 3.5 (b), (d) for $p$-polarization (a), (b) and $s$-polarization (c), (d). $F$ (black solid curve) is the binding force while $F_{e,r}$ (red solid curve), $F_m$ (blue solid curve), and $F_{em}$ (green solid curve) are electric, magnetic, and coupling components of $F$. Additionally, $F_{Ray}$ (purple dashed curve) is the binding force on the same sphere in electric dipolar dimer calculated by the typical Rayleigh approximation [Eq. (3b) in Ref. [15]]. The digits in (a), (c) and (d) indicates a multiple of the magnified force. The solid red triangles, blue stars, green diamonds, black circles, and hollow purple circles represent respectively the stable equilibrium positions of $F_e$, $F_m$, $F_{em}$, $F$, and $F_{Ray}$.

## Appendix H: THE DIFFERENCE BETWEEN $F_e$ AND $F_{Ray}$ FOR $P$- AND $S$-POLARIZATION

Figure S2 shows $F_e$ together with each of its terms and $F_{Ray}$ as function of distance ($R$) between the centers of two spheres with $n_p$=3.5 for $p$-polarization (a) and $s$-polarization (b). Figure S2 (a) shows that $F_{Ray}$ overestimates (underestimates) $F_e$ at the first dip (peak) of $F_e$ in front of the first stable equilibrium position of $F$ for $p$-polarization. From physical viewpoint, in the vicinity of the dip, the phases of the two $p_y$ in the different particles, which are respectively induced by the incident electric field and the re-radiated electric field by the electric dipole in the neighboring particle, are almost synchronous. And then, the two $p_y$ form a head-to-tail colinear configuration and attract each other, the same as the Rayleigh dimer. In addition, the hybridization arises because of the increase of the refractive index of particles. In detail, the $m_x$ in one particle induced by the incident magnetic field excites $p_z$ in the neighboring one. Figure S2 (a) shows that the force due to the interaction between $p_y^A$ and $p_y^B$ (red dashed curve), which is described by the first term in Eq. (1) in the main text, coincides with $F_{Ray}$ (purple thick dash-dotted curve) because of the same physical origin. Additionally, the interaction between $p_z^A$ and $m_x^B$ is repulsive (green dash-dot-dotted curve) while $p_z^A$ and $p_z^B$ are antiparallel and attractive (blue short dashed curve). The two forces are respectively expressed by the third and second terms in Eq. (1) in the main text. But the latter is the secondary interaction and smaller than the former. Therefore the attractive force is partly offset by the repulsive force. It means that the $F_e$ (black solid curve) is reduced by the hybridization compared to $F_{Ray}$. In a word, $F_{Ray}$ overestimates $F_e$ at the dip of $F_e$. On the contrary, at the peak of $F_e$ mentioned above, the phases of the two $p_y$ are nearly inverse since the increase of $R$. This leads to head-to-head colinear configuration of the two $p_y$. Thus the force due to the interaction between $p_y^A$ and $p_y^B$ (red dashed curve) becomes repulsive while the interaction between $p_z^A$ and $m_x^B$ is still repulsive (green dash-dot-dotted curve). As a result, the repulsive $F_e$ is enhanced which means that $F_{Ray}$ underestimates $F_e$ at the peak of $F_e$. But the difference between $F_e$ and $F_{Ray}$ is unremarkable at large $R$ because of the weak hybridization.

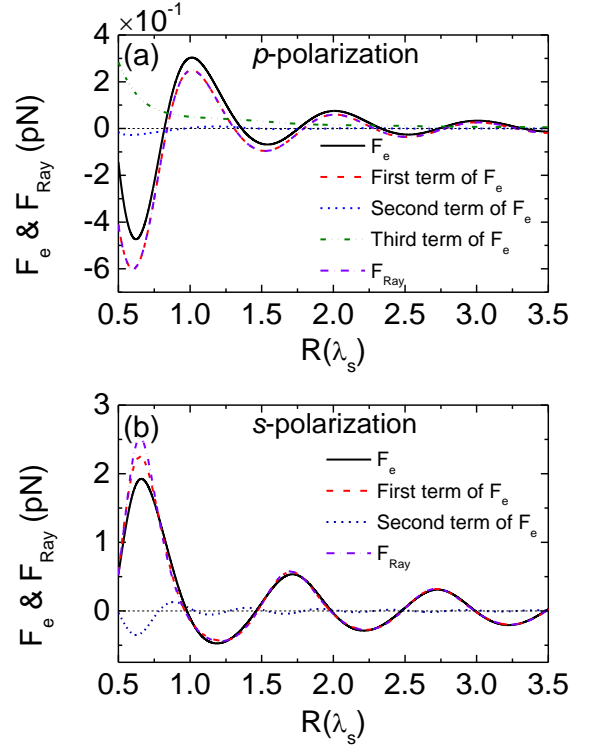

Fig. S2. $F_e$ together with its all terms and $F_{Ray}$ versus distance ($R$) between the centers of two spheres with refractive index $n_p$=3.5 for $p$-polarization (a) and $s$-polarization (b). (a) $F_e$ (black solid curve) and $F_{Ray}$ (purple thick dash-doted curve) are the same as the counterparts in Fig. S1 (b). The red dashed, blue short dashed, and green dash-dot-dotted curves are respectively calculated by the first, second, and third terms in Eq. (1) in the main text. Meanwhile, they denote respectively the interactions between $p_y^A$ and $p_y^B$, between $p_z^A$ and $p_z^B$, and between $p_z^A$ and $m_x^B$. (b) $F_e$ and $F_{Ray}$ are the same as the corresponding forces in Fig. S1 (d). The red dashed and blue short dashed curves are individually calculated by the fourth and fifth terms in Eq. (1) in the main text. The former denotes the interaction between $p_x^A$ and $p_x^B$ while the latter indicates the interaction between $p_x^A$ and $m_z^B$. The parameters are the same as in Fig. 2 in the main text.

It is contrary to the $p$-polarization that Fig. S2 (b) denotes that $F_{Ray}$ underestimates (overestimates) $F_e$ at the first dip

(peak) of $F_e$ in the case of *s*-polarization. This can be also explained by the hybridizations as in the analyses of *p*-polarization. In detail, the re-radiated magnetic field by $p_x$ in one particle is strong enough to induce $m_z$ in the neighboring one. On the one hand, the interaction between $p_x^A$ and $p_x^B$ expressed by the fourth term in Eq. (1) in the main text is repulsive (red dashed curve). On the other, the interaction between $p_x^A$ and $m_z^B$ (blue short dashed curve) described by the fifth term in Eq. (1) in the main text is attractive in front of the first peak of $F_e$. And then the former is partly offset by the latter, which is the reason why $F_{Ray}$ (purple thick dash-doted curve) is larger than $F_e$ (black solid curve) around the first peaks of $F_e$. However, the interaction between $p_x^A$ and $m_z^B$ is still attractive when the interaction between $p_x^A$ and $p_x^B$ changes to attraction at the first dip of $F_e$. As a result, $F_e$ is enhanced and larger than $F_{Ray}$. As expected, the effects become weaker and weaker with increment of distance ($R$).

**APPENDIX I: STABLE EQUILIBRIUM POSITIONS OF BINDING FORCE IN *P*- AND *S*- POLARIZATIONS.**

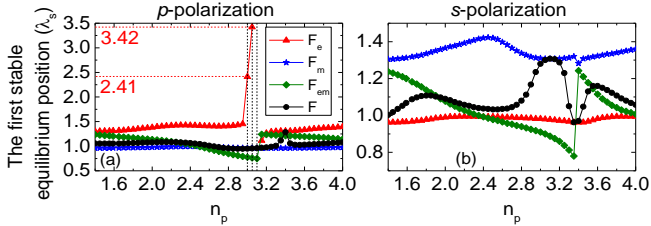

Fig. S3. The first stable equilibrium positions of $F$ (black curve with circles), $F_e$ (red curve with triangles), $F_m$ (blue curve with stars) and $F_{em}$ (green curve with diamonds) versus refractive index ($n_p$) of the particles for *p*-polarization (a) and *s*-polarizations (b). The particle radius is 150nm. The two horizontal red dashed lines denote the positions $2.41\lambda_s$ and $3.42\lambda_s$. The three vertical black dashed lines in (a) correspond to the locations $n_p$ =3, 3.05, and 3.1.

We have calculated the first stable equilibrium positions of $F$, $F_e$, $F_m$, and $F_{em}$ for *p*- and *s*- polarizations. At these positions, the particles experience a restoring force whose intensity is zero and slope is negative. Importantly, the stability of the dimer can be characterized by the stiffness defined as the absolute of the slope of restoring force at the stable equilibrium positions [44]. In general, it is only interesting in the first stable equilibrium position of the binding force because the bound state of the dimer is the most robust at the position due to the maximum stiffness. For *p*-polarization, the first stable equilibrium position of $F_e$ (red curve with triangles) in Fig. S3 (a) fluctuates slightly around $1.3\lambda_s$ with increase of $n_p$. However, the position moves rapidly to $2.41\lambda_s$ and $3.42\lambda_s$ (two horizontal red dashed lines) at locations $n_p$ =3 and 3.05 (left and middle vertical black dashed lines), respectively. Meanwhile, the stiffness of $F_e$ decreases sharply at the two locations, meaning $F_e$ is hardly sufficient to enable the dimer to form a stable bound state. Moreover, the stable equilibrium even vanishes at $n_p$ =3.1 (right vertical black dashed line) where $F_e$ has no stiffness (vertical black dashed line). As expected, $F_e$ without stable equilibrium cannot cause the dimer to form stable configuration at all (see details in Appendix J). On the other hand, the first stable equilibrium position of $F_{em}$ (green curve with diamonds) fluctuates with increase of $n_p$. Importantly, the first stable equilibrium position of $F$ (black curve with circles) is close to the counterpart of $F_m$ (blue curve with stars) around $\lambda_s$ for all $n_p$. For *s*-polarization in Fig. S3 (b), the stable equilibrium position of $F$ is close to that of $F_e$ and $F_{em}$ in the regions $1.4< n_p <2.9$ and $2.5< n_p <4$.

**APPENDIX J: $F$ AND $F_e$ ON PARTICLE WITH $n_p$=3, 3.05, AND 3.1 FOR *P*-POLARIZATION**

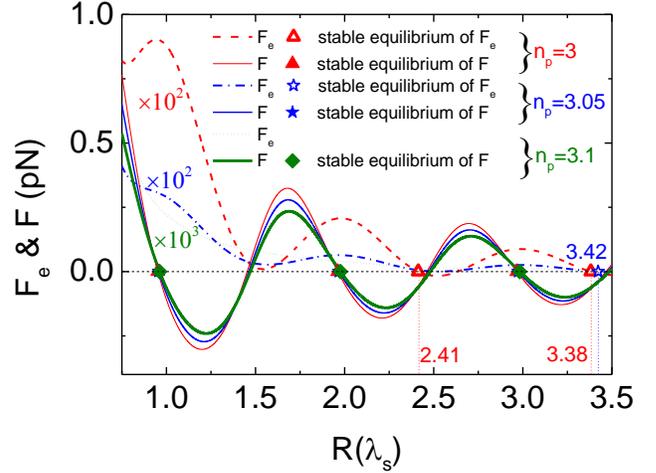

Fig. S4. Binding force $F$ calculated by the sum of Eqs. (1)-(3) in the main text and the electric component $F_e$ in Eq. (1) versus distance ($R$) between the centers of two spheres with refractive index $n_p$=3 (red), 3.05 (blue) and 3.1 (green) for *p*-polarized wave. The solid thin red, blue, and thick green curves represent individually $F$ for $n_p$=3, 3.05, and 3.1. The digits indicate a multiple of the magnified forces. The solid red triangles, blue stars, and green diamonds represent respectively the stable equilibrium positions of $F$ while the hollow red triangles and blue star denote the counterparts of $F_e$ for $n_p$=3, 3.05, and 3.1. The hollow red triangles at positions $R$=$2.41\lambda_s$ and $3.38\lambda_s$ denote the first and second stable equilibrium positions of $F_e$ on particles with $n_p$=3. The hollow blue star at position $R$=$3.42\lambda_s$ denotes the unique stable equilibrium position of $F_e$ for particles with $n_p$=3.05. The solid thick green short-dotted curve corresponding to $F_e$ for $n_p$=3.1 has no stable equilibrium position. The radius of particle and parameters of both water and wave are same as in Fig. 2 in the main text.

Figure S4 shows $F$ calculated by the sum of Eqs. (1)-(3) in the main text and the electric component $F_e$ in Eq. (1) versus distance ($R$) between the centers of two spheres with refractive index $n_p$=3 (red), 3.05 (blue) and 3.1 (green) for a *p*-polarized wave. It can be seen that $F_e$ (red dashed and blue dash-dotted curves) has respectively two (red hollow triangle at $2.41\lambda_s$ and $3.38\lambda_s$) and one (blue hollow star at $3.42\lambda_s$) stable equilibrium positions for particles with $n_p$=3 and 3.05. Meanwhile, $F_e$ is two order of magnitude less than $F$ (red and blue solid curves) for both cases of $n_p$=3 and 3.05. In addition, the stiffness of $F_{em}$ is also far less than that of $F$ (see Fig. 3 (a) in the main text). As a result, $F$ is dominated by $F_m$. The stable equilibrium positions of $F$ are respectively shown by the solid red triangles and blue stars for $n_p$=3 and 3.05. Moreover, for particles with $n_p$=3.1, $F_e$ (green short-dotted curve) is three orders of magnitude less than $F$

(green solid curve) and has no stable equilibrium position while the stable equilibrium positions of $F$ (solid green diamonds) are dominated by $F_m$.

## APPENDIX K: RADIALLY STABLE EQUILIBRIUM POSITIONS IN CASE OF CIRCULAR POLARIZATION.

In the case of circular polarization, the first radially stable equilibrium position of $F_{em}$ (green curve with solid diamonds) in Fig. S5 fluctuates with increase of $n_p$ while that of $F_e$ (red curve with solid triangles) jumps in the vicinity of $n_p$=3.1. Interestingly, the corresponding positions of $F_m$ (blue curve with solid stars) and $F$ (black curve with solid circles) are near $\lambda_s$ and change little. As expected, other stable equilibrium positions of $F$ lie in near-integer multiples of $\lambda_s$ while the radial stability of the dimer at the positions decreases in turn.

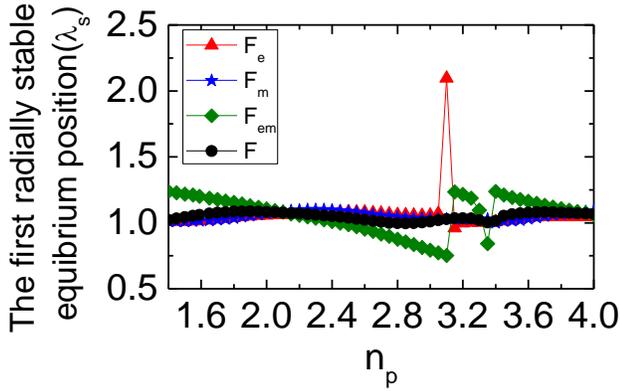

Fig. S5. The first radially stable equilibrium positions of the binding force ($F$) including all three components ($F_e$, $F_m$, and $F_{em}$) versus refractive index ($n_p$) of particles for left-circularly polarized wave. The parameters are same as in Fig. 2 in the main text.

## APPENDIX L: BINDING FORCE AND TORQUE FOR LOW- AND MODERATE-REFRACTIVE-INDEX PARTICLES FOR CIRCULAR POLARIZATION.

Figures S6 (a) and (b) show $F_e$, $F_m$, $F_{em}$, and $F$ with $n_p$ =1.59 and 3.5 for circularly polarized wave. Firstly, for low-refractive-index in (a), the electric dipole has advantage over the magnetic dipole. And then, $F_e$ (red dashed curve) is greatly larger than $F_m$ (blue dash-dot-doted curve) and $F_{em}$ (green short dashed curve). As a result, $F$ (black solid curve) is mainly dominated by $F_e$ while the stable equilibrium positions of $F$ (black solid circles) are very close to the counterparts of $F_e$ (not marked). This is similar to the $p$- and $s$-polarized cases in Figs. S1 (a) and (c). Secondly, for moderate-refractive-index in (b), the electric and magnetic dipoles are effectively induced. Therefore, all components of $F$ are enhanced by two orders of magnitude compared to Fig. S6 (a). Naturally, $F_e$, $F_m$ and $F_{em}$ all contribute non-negligibly to $F$. As expected, the stable equilibrium positions of $F$ are determined by the $F_e$, $F_m$, and $F_{em}$ together.

Figures S6 (c) and (d) show $\Gamma_e$, $\Gamma_m$, $\Gamma_{em}$, and $\Gamma$ with $n_p$ =1.59 and 3.5 for circularly polarized wave. Firstly, even though for particles with low refractive index in Fig. S6 (c), $\Gamma$ (black solid curve) is almost dominated by $\Gamma_{em}$ (green short dashed curve) since $\Gamma_m$ (blue dash-dot-dotted curve) and $\Gamma_e$ (red dashed curve) are respectively twenty and five times smaller than $\Gamma_{em}$. Meanwhile, $\Gamma_m$ and $\Gamma_e$ decay with increase of $R$. Interestingly, the envelopes of $\Gamma_{em}$ and $\Gamma$ do not decay as shown by Eqs. (11) and (12) in the main text. Secondly, for particles with moderate refractive index in Fig. S6 (d), $\Gamma$ and $\Gamma_{em}$ are dramatically strengthened by three orders of magnitude compared to Fig. S6 (c). It means that the increase of refractive index of particles is in favor of torque. In addition, $\Gamma_m$ is largely enhanced relative to $\Gamma$ compared to Fig. S6 (c). The two facts benefit from the enhanced magnetic response and hybridization of the two particles. Here, $\Gamma$ is still controlled by $\Gamma_{em}$ because $\Gamma_{em}$ is stronger than $\Gamma_e$ and $\Gamma_m$.

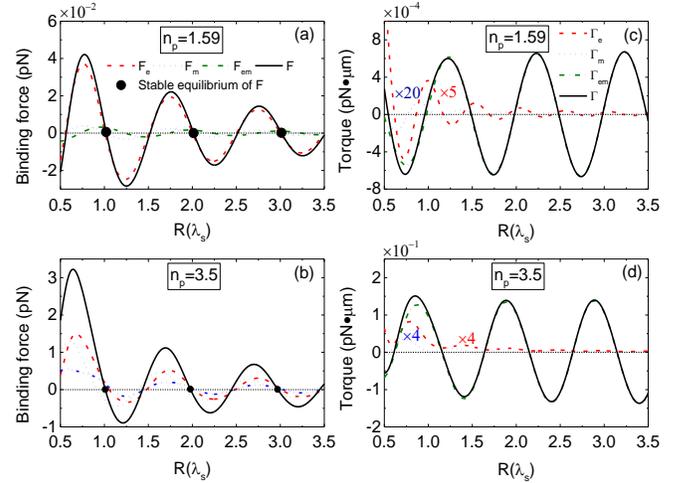

Fig. S6. Binding force and torque on particles for a left-circularly polarized wave. Total binding force ($F$) and each of the three components ($F_e$, $F_m$, and $F_{em}$) on sphere B versus distance ($R$), with refractive index $n_p$=1.59 (a) and 3.5(b). The solid black circles denote the stable equilibrium positions of $F$. Total torque ($\Gamma$) upon the dimer and each of the three components ($\Gamma_e$, $\Gamma_m$, and $\Gamma_{em}$) versus distance ($R$) with refractive index $n_p$=1.59 (c) and 3.5(d). The digits in (c) and (d) indicate a multiple of the magnified torques. The parameters are the same as in Fig. 2 in the main text.